\documentstyle [prl,aps,twocolumn,epsfig]{revtex}
\newcommand{\la}[1]{\label{#1}}

\newcommand{\A}{{\bf A}}
\newcommand{\F}{{\bf F}}
\newcommand{\B}{{\bf B}}
\renewcommand{\d}{{\bf d}}

\newcommand{\n}{{\bf n}}

\newcommand{\be}{\begin{equation}}
\newcommand{\ee}{\end{equation}}
\newcommand{\ba}{\begin{eqnarray}}
\newcommand{\ea}{\end{eqnarray}}
\newcommand{\bastar}{\begin{eqnarray*}}
\newcommand{\eastar}{\end{eqnarray*}}

\newskip\humongous \humongous=0pt plus 1000pt minus 1000pt

\newif\ifdtup

\relax

\begin{document}
%
\title  {Partially Dual Variables in SU(2) Yang-Mills Theory
}
\bigskip

\author{Ludvig Faddeev$^{* \sharp}$ and Antti J. Niemi$^{** \sharp}$} 

\address{
$^*$St.Petersburg Branch of Steklov Mathematical
Institute, 
Russian Academy  of Sciences, Fontanka 27 , St.Petersburg, 
Russia \\ $^{**}$Department of Theoretical Physics,
Uppsala University, 
P.O. Box 803, S-75108, Uppsala, Sweden
\\$^{\sharp}$Helsinki Institute of Physics, 
P.O. Box 9, FIN-00014 University of Helsinki, Finland \\
{\scriptsize \bf FADDEEV@PDMI.RAS.RU, NIEMI@TEORFYS.UU.SE} \\ \vskip 0.3cm
}

\maketitle

\begin{abstract}
We propose a reformulation of SU(2) Yang-Mills theory in 
terms of new variables. These variables are 
appropriate for describing the theory in its infrared limit, and  
indicate that it admits knotlike 
configurations as stable solitons. As a consequence
we arrive at a dual picture of the Yang-Mills theory 
where the short distance limit describes asymptotically 
free, massless point gluons and the large distance limit 
describes extended, massive knotlike solitons. 

\end{abstract}

\narrowtext
\bigskip

In the high energy limit Yang-Mills theory is 
asymptotically free, and can be solved perturbatively.
It describes the interactions of massless gluons which 
correspond to the transverse polarizations of the 
gauge field $A_\mu$ \cite{ymthy}.

At low energies Yang-Mills theory becomes strongly coupled.
Perturbative techniques fail and nonperturbative methods
must be developed. For this substantial efforts
have been devoted, but numerical lattice 
approaches still remain the most viable tool 
to effectively explore the low 
energy theory. But in spite of our 
lacking theoretical understanding of low energy
Yang-Mills theory, we expect that it exhibits 
color confinement with ensuing mass gap. The physical spectrum 
is supposed to describe massive composites of $A_\mu$ 
such as glueballs. When quarks are introduced the 
gauge field should form string-like flux 
tubes which confine quarks inside hadrons.

In the present Letter we propose an approach 
to investigate SU(2) Yang-Mills theory in the infrared 
limit. Our proposal is motivated by the 
qualitative picture developed in particular by 'tHooft
and Polyakov \cite{thooft}, who asserted that the ultraviolet
and infrared limits of a Yang-Mills theory 
represent different phases, with color confinement due 
to a dual Meissner effect in a condensate of
magnetic monopoles. This picture suggests that even though 
the gauge field $A_\mu$ is the proper order parameter 
for describing the theory in its ultraviolet limit, in 
the infrared limit with monopole condensation
some other order parameter could become more adequate.
Naturally we expect, that such a change of variables may   
also imply certain need to reformulate the Yang-Mills action.

In the high energy limit the theory is described by 
the standard Yang-Mills action
\be
S \ = \ \frac{1}{g^2}\int dx \ Tr F^2
\la{YM1}
\ee
This is the {\it unique} Lorentz and gauge invariant local
action which is renormalizable in four dimensions and admits
a Hamiltonian interpretation which identifies the transverse 
polarizations of $A_\mu$ as the physical fields
present in the ultraviolet limit.

In the following we shall propose new variables
for describing the infrared limit of a four dimensional
SU(2) Yang-Mills theory. We shall argue that instead of
$A_\mu$, in this limit the appropriate order parameter 
is a three component vector $n^a(x) \ (a = 1,2,3)$
with unit length $\n \cdot \n = 1$ and
classical action \cite{fadde}
\be
S \ = \ \int dx \ m^2 (\partial_\mu \n)^2 \ + \
\frac{1}{e^2} ( \n , \d\n \times \d\n)^2 
\la{fad1}
\ee
Here $m$ is a mass scale and $e$ is a dimensionless 
coupling constant. This is the {\it unique} 
local and Lorentz-invariant action for the unit 
vector $\n$ which is at most quadratic in time 
derivatives so that it admits a Hamiltonian 
interpretation, and involves {\it all} such terms 
that are either relevant or marginal in the 
infrared limit. 

Observe that the action (\ref{fad1}) can 
be related to the $SU(2)$ Skyrme model,
restricted to a sphere $S^2$. However,
the topological features of these two 
models are quite different.

We shall argue that (\ref{fad1}) emerges from (\ref{YM1}) 
by a change of variables together with a renormalization
group argument. Thus it can be considered as a unique 
action for describing the low energy limit of 
a SU(2) Yang-Mills theory, 
in par with its high energy limit counterpart
(\ref{YM1}).

We note that in four dimensions the action (\ref{fad1})
fails to be perturbatively renormalizable in the
ultraviolet. But since it is expected to describe
the physical excitations of a SU(2) Yang-Mills theory
only in the low energy strong coupling limit, lack
of perturbative renormalizability should 
not pose a problem provided we can interpret
(\ref{fad1}) adequately: In the following
we shall argue that (\ref{fad1}) can be derived 
from (\ref{YM1}) by a renormalization group
improved change of variables. Since (\ref{YM1})
is renormalizable, this suggests that the quantum theory
of (\ref{fad1}) should also be consistent when
properly treated. Indeed, we have 
recently established \cite{nature} that in 3+1 dimensions the 
classical action (\ref{fad1}) describes stable 
knotlike solitons. This suggests that a proper 
route to its quantization should be based on the 
investigation of the quantum mechanical properties of 
these solitons. 

From the point of view of a Yang-Mills 
theory the presence of knotlike solitons is actually  
quite appealing. It is natural to relate these 
solitons with the string-like flux tubes that we
expect to be present in the infrared spectrum of a
Yang-Mills theory, to provide the confining force 
between two quarks. In the absence of quarks such 
flux tubes may still be present as color-neutral
excitations. They now close on themselves in knotted, 
stable solitonic configurations which are natural 
candidates for describing glueballs. In this manner 
we arrive at a dual picture of the Yang-Mills theory, with 
the high energy limit described by massless and
pointlike transverse polarizations of $A_\mu$ and
the low energy limit described by massive
solitonic flux tubes which close on themselves in
stable knotlike configurations

We shall now proceed to justify the action
(\ref{fad1}). We are motivated by the picture 
developed in \cite{thooft}, with confinement viewed  
as a dual Meissner effect in a condensate of magnetic monopoles. 
In a SU(2) Yang-Mills theory the relevant magnetic monopole is 
the (singular) Wu-Yang configuration \cite{wu}
\be
A_i^a \ = \ \epsilon_{aik}\frac{x_k}{r^2}
\la{wu1}
\ee
and in order to describe a condensate of these
monopoles, we need to properly extend 
(\ref{wu1}) by introducing a smooth field for the
corresponding order parameter.  A natural
Ansatz for extending (\ref{wu1}) into a condensate is  
\be
A_i^a \ = \ \epsilon_{abc} \partial_i n^b n^c \
\equiv \ \d\n \times \n 
\la{wu2}
\ee  
with $\n$ a three component unit vector field
that describes the condensate. It 
reproduces (\ref{wu1}) when we specify to the 
singular "hedgehog" configuration 
$\n = {\bf x}/ r$.  

The unit vector $\n$ describes two independent 
field variables. Since a gauge fixed four 
dimensional SU(2) connection
$\A_\mu$ describes six polarization degrees of freedom,  
we need to extend the parametrization (\ref{wu2}) by
four additional polarizations. In order to 
search for a natural extension, we first observe 
that under an infinitesimal gauge transformation
\[
\delta A^a_\mu \ = \ \nabla^{ab}_\mu \varepsilon ^b 
\ = \ \partial_\mu \varepsilon^a
+  \epsilon^{acb}A^c_\mu \varepsilon^b
\]
which is parametrized by the Lie algebra 
element $\varepsilon^a(x) = \varepsilon(x) \cdot 
n^a(x)$, (\ref{wu2}) fails to 
remain form invariant. But if we improve (\ref{wu2}) into
\be
\A_\mu  \ = \ C_\mu \n  \ + \ \d \n \times \n
\la{wu3}
\ee
where $C_\mu(x)$ is a vector field which transforms as 
an abelian connection 
\be
C_\mu \ \to \ C_\mu + \partial_\mu \varepsilon
\la{gi}
\ee
the functional form of the configuration (\ref{wu3}) 
remains intact under this gauge transformation. 

The functional form (\ref{wu3}) 
of SU(2) connections has been previously studied in particular 
by Cho \cite{cho}, as a consistent truncation of the 
full four dimensional connection $A^a_\mu$. 
The goal in his work is to identify those field degrees of freedom 
in $A^a_\mu$ which are 
relevant for describing the Abelian dominance, 
a concept that originates from \cite{thooft} and is 
expected to be relevant for color confinement. 

The abelian gauge invariance (\ref{gi}) implies that (\ref{wu3})
describes four field components, corresponding
to the two transverse polarizations of the U(1) 
connection $C_\mu$
and the two independent components of $\n$.
In order to extend (\ref{wu3}) so that it
describes all six field components of
an arbitrary connection $\A_\mu$, we consider
an arbitrary finite gauge transformation of a generic
connection $A_\mu$. With
\[
U(x) \ = \ \exp \{ i \frac{1}{2} \alpha \n \cdot \tau \}
\]
the SU(2) group element that determines this
gauge transformation, we find for the 
gauge transformation of an arbitrary
connection $A_\mu^a$
\[
\A^U \ = \ [(\A,\n) + \d \alpha] \n \ + \ \d\n \times 
\n 
\]
\be
+ \ \sin \ \alpha \cdot ( \d\n 
+ \A \times \n) -
\cos \ \alpha \cdot(\d\n + 
\A \times \n)\times \n 
\la{gt}
\ee
From this we conclude that  a
generic connection $\A_\mu$ 
should have the functional form 
\be
\A_\mu \ = \ C_\mu \n \ + \ \d\n \times \n \ + \ 
\varphi \B_\mu \ + \ \B_\mu \times  \n
\la{wu4}
\ee
where $\varphi$ is a scalar field and $B^a_\mu$
is an orthogonal SU(2) valued 
vector, $\n \cdot \B_\mu = 0$ for all $\mu$.
Since the number of independent field components
carried by a four dimensional SU(2) connection
is six, the orthogonal field $\B_\mu$ should
only describe a single component, 
and we can select
it to be proportional to $\d\n$. This yields the 
following Ansatz for parametrizing a generic four 
dimensional connection,
\be
\A_\mu \ = \ C_\mu \n \ + \ \d\n \times \n \ + \ \rho  
\d \n \ + \ \sigma \d\n \times \n 
\la{wu5}
\ee
Notice that we have here separated the
second and fourth terms in the {\it r.h.s.}, even though
these terms are linearly dependent.  
The reason for this separation is, that it allows us to 
combine the scalars $\rho$ and $\sigma$ into a complex field
\be 
\phi = \rho + i \sigma
\la{phi}
\ee
with the property that under a SU(2) gauge transformation
generated by $\alpha^a = \alpha \cdot \n$ the functional form
of (\ref{wu5}) remains intact, with the multiplet 
$(C_\mu, \phi )$ transforming like the field multiplet in
the abelian Higgs model.  

In order to verify that our parametrization (\ref{wu5}) is
indeed complete, we substitute it to the classical Yang-Mills
action (\ref{YM1}) and derive equations of motion obtained 
by varying the component fields $(\n , C_\mu, \phi)$. These 
equations should reproduce the original Yang-Mills equations, 
obtained by {\it first} varying {\it w.r.t.}
$\A_\mu$ in (\ref{YM1}) and then substituting 
(\ref{wu5}):

If we introduce the U(1) covariant derivative
\[
D_\mu \phi \ = \ \partial_\mu \phi \ + \ i C_\mu \phi 
\]
\be
\ = \ \partial_\mu \rho \ - \ C_\mu \sigma \ + \ 
i \ ( \partial_\mu \sigma \ + \ C_\mu \rho )
\ = \  {D}_\mu \rho \ + \ i {D}_\mu \sigma
\la{covder}
\ee
we find 
\[
\F_{\mu\nu}  =  \n ( G_{\mu\nu} - [ 
1 - (\rho^2 + \sigma^2) ] H_{\mu\nu})
+  ( {D}_\mu \rho \ \partial_\nu \n 
- {D}_\nu \rho \ \partial_\mu \n ) \\ 
\]
\be
 +  \ ( {D}_\mu \sigma \ \partial_\nu \n \times \n
\ - \ {D}_\nu \sigma \ \partial_\mu \n \times \n )
\la{F1}
\ee
where
\bastar
G_{\mu\nu} \ & = & \ \partial_\nu C_\mu - \partial_\mu C_\nu \\
H_{\mu\nu} \ & = & \ ( \n \ , \ \partial_\mu 
\n \times \partial_\nu \n ) \\
\eastar
When we substitute (\ref{F1}) into the Yang-Mills action
(\ref{YM1}) we get
\[
S   =  \frac{1}{g^2}\int dx  \biggl\{ \ \n \left[ 
F_{\mu\nu} 
- ( 1 - [\rho^2 + \sigma^2 ]) H_{\mu\nu}
\right]  +  ( D_\mu \rho \ \partial_\nu \n  \\
\]
\be
- D_\nu \rho \ \partial_\mu \n ) \
+  \ (D_\mu \sigma \partial_\nu \n \times \n 
- D_\nu \sigma \partial_\mu \n
\times \n ) \biggr\}^2
\la{YM2}
\ee
and when we perform the variations {\it w.r.t.}
$(C_\mu, \phi, \n)$ we get
\bastar
\n \cdot \nabla_\mu \F_{\mu\nu} \ & = & \ 0 \\
\partial_\nu \n \cdot \nabla_\mu \F_{\mu\nu} \ & = & \ 0 \\
\partial_\nu \n \times \n \cdot \nabla_\mu \F_{\mu\nu}
\ & = & \ 0 \\
\left( \ D_\nu \rho \ + \ D_\nu \sigma \cdot \n \times 
\ \right) \cdot \nabla_\mu \F_{\mu\nu} \ & = & \ 0 
\eastar
which are all proportional to 
the ordinary Yang-Mills equation, evaluated at the field
(\ref{wu5}).
But the U(1) invariance (\ref{gi}) implies that
only six of these equations can be independent. These
equations coincide with the six independent second 
order equations that we obtain when we first vary 
the action (\ref{YM1}) {\it w.r.t.} the full 
connection $A_\mu^a$ and then substitute for the 
parametrization (\ref{wu5}). Thus we assert that the
parametrization (\ref{wu5}) is indeed complete.  
(We remind that the variation of (\ref{YM1}) {\it
w.r.t.} $A_\mu^a$ yields twelve equations, but the three
$A_0^a$ are Lagrange multipliers and three of the equations
are first order, corresponding to Gauss law in the Hamiltonian
approach. Consequently in four dimensional SU(2) Yang-Mills
theory there are only six independent second order equations.)

We observe that the second term in (\ref{fad1}) 
is also present in (\ref{YM2}). Since the first term in 
(\ref{fad1}) involves a mass scale it is absent in (\ref{YM1}), 
as there is no way to introduce a mass scale in four
dimensional Yang-Mills by employing ultraviolet 
renormalizable, local, Lorentz and gauge 
invariant functionals of $A_\mu$. However, when we
represent the Yang-Mills action using the component field
(\ref{wu4}), (\ref{wu5}), there is nothing {\it
a priori} that would prevent us from including the first term in
(\ref{fad1}) already at the tree level. Indeed, since 
it is a relevant operator in the infrared, even
if absent at the tree level it should emerge when we account 
for quantum fluctuations in a gradient expansion. 
We assert that these fluctuations produce 
a non-vanishing expectation value when we average over the
scalar field $\phi = \rho + i \sigma$ in (\ref{YM2}) 
to the effect 
\be
<|\partial_\lambda \phi|^2  \ \eta_{\mu\nu}
- \partial_\mu \phi^* \partial_\nu \phi> \ = \ m^2 \eta_{\mu\nu}
\la{aver}
\ee
As a consequence we conclude that 
the full action (\ref{fad1}) is
contained in a gradient expansion of 
the effective action for the order parameter $\n$. 

Obviously the full effective action for the order 
parameter $\n$ obtained by integrating over the complete  
set of fields in the parametrization (\ref{wu4}), will  
also contain various additional functionals of $\n$ 
besides the two terms that appear in (\ref{fad1}). 
However, (\ref{fad1}) is {\it unique} 
in the sense that it contains {\it all} such infrared 
relevant and marginal, local Lorentz invariant operators 
of $\n$ which are at most quadratic in time derivatives,
as is necessary for a Hamiltonian interpretation. 
Consequently we may as well
adopt the point of view, that
(\ref{fad1}) is the {\it unique} fundamental action 
to describe the low energy limit of a SU(2) 
Yang-Mills theory, in the confining phase where magnetic monopoles
condense. There is no alternative which would
be consistent with our general principles! 
The results of \cite{nature} then suggest 
that at low energies the physical states of the Yang-Mills
theory are knotlike solitons of the monopole condensate,
and it becomes natural to view these configurations as candidates 
for describing glueballs.

Notice that the present interpretation of (\ref{fad1})
is entirely analogous to the common 
point of view to consider (\ref{YM1}) as the fundamental 
action for the high energy Yang-Mills theory, even 
though {\it e.g.} a gradient expansion of the
lattice Yang-Mills action involves higher derivative
terms which all become irrelevant in the 
continuum (short-distance) limit 
where the lattice spacing tends to zero. 

Besides the order parameter $\n$ which is appropriate for
describing the phase with monopole condensation, we have 
also found that the abelian Higgs multiplet $(C_\mu , \phi)$ 
naturally appears in the parametrization
of four dimensional connections. 
Elimination of $\n$ in (\ref{YM2}) 
then produces an effective action for the
abelian Higgs multiplet which comprises a natural 
order parameter for describing the SU(2) 
theory in a "Higgs phase", also considered 
in \cite{thooft}. Indeed, since 
we have the spontaneously broken Higgs self-coupling 
present in (\ref{YM2}),
\[
V(\phi) \ = \ <H_{\mu\nu}^2> \cdot (1 -  [\rho^2 
+ \sigma^2])^2
\ \sim \ \lambda ( 1 - |\phi |^2 )^2
\]
we can expect the corresponding effective action
to support Nielsen-Olesen type 
vortices as infinite energy line 
solitons. In a sense, these abelian Higgs variables
can be viewed as dual to the vector field
$\n$ in the expansion (\ref{wu5}), and
it would be of interest
to further study the properties of this "Higgs phase".

\vskip 0.4cm

In conclusion, we have argued that (\ref{fad1})
is the unique action for describing SU(2) 
Yang-Mills theory at low energies, consistent
with various natural first principles. 
It can be derived directly from (\ref{YM1}) by
a renormalization group improved change of variables.  
The unit vector $\n$ can be viewed as an order  
parameter for monopole condensation, and the first term
in (\ref{fad1}) should be included since
it is relevant in the infrared limit. This term 
introduces a mass gap and the ensuing 
action supports knotlike configurations as stable 
solitons. This suggests a dual picture of the 
Yang-Mills theory where the high energy limit 
describes massless pointlike gluons and 
the infrared limit describes massive knotted solitons,
consistent with the commonly accepted picture of 
color confinement. Furthermore, we have
found that the parametrization
of a generic connection also contains the abelian Higgs 
multiplet, in a manner which can be 
viewed as dual to the vector field $\n$. 
This suggests that line vortices of the 
abelian Higgs model may also be present in a 
description of the theory in an appropriate 
"Higgs phase", in line with the original picture
in \cite{thooft}. It would be interesting to
further study the properties of this phase.

{\bf Aknowledgements} L.F. aknowledges a stimulating 
discussion with S.Shabanov, and we are both indebted 
to Y.M. Cho for discussions. The research by L.F. was 
partially supported by Russian Academy of Sciences and 
Russian Foundation for Fundamental Research Grant 
RFFR-96-01-00851, and the research by A.N. was 
partially supported by NFR Grant F-AA/FU 06821-308.

\vfill
                      
\eject

\end{document}